\RequirePackage{ifpdf}
\ifpdf 
\documentclass[pdftex]{sigma}
\else
\documentclass{sigma}
\fi

\numberwithin{equation}{section}

\begin{document}

\allowdisplaybreaks

\renewcommand{\thefootnote}{$\star$}

\renewcommand{\PaperNumber}{096}

\FirstPageHeading

\ShortArticleName{Supersymmetric Extension of Non-Hermitian su(2)
Hamiltonian and Supercoherent States}

\ArticleName{Supersymmetric Extension of Non-Hermitian\\ su(2)
Hamiltonian and Supercoherent States\footnote{This
paper is a contribution to the Proceedings of the Workshop ``Supersymmetric Quantum Mechanics and Spectral Design'' (July 18--30, 2010, Benasque, Spain). The full collection
is available at
\href{http://www.emis.de/journals/SIGMA/SUSYQM2010.html}{http://www.emis.de/journals/SIGMA/SUSYQM2010.html}}}

\Author{Omar CHERBAL~$^\dag$, Mahrez DRIR~$^\dag$, Mustapha MAAMACHE~$^\ddag$ and Dimitar A.~TRIFONOV~$^\S$}

\AuthorNameForHeading{O.~Cherbal, M.~Drir, M.~Maamache and D.A.~Trifonov}

\Address{$^\dag$~Faculty of Physics, Theoretical Physics Laboratory,
USTHB, \\
\hphantom{$^\dag$}~B.P. 32, El Alia, Algiers 16111, Algeria}
\EmailD{\href{mailto:ocherbal@yahoo.fr}{ocherbal@yahoo.fr}, \href{mailto:ocherbal@usthb.dz}{ocherbal@usthb.dz}}

\Address{$^\ddag$~Laboratoire de Physique Quantique et Systemes
Dynamiques, Department of Physics,\\
\hphantom{$^\ddag$}~Setif University, Setif 19000, Algeria}
\EmailD{\href{mailto:maamache_m@yahoo.fr}{maamache\_m@yahoo.fr}}

\Address{$^\S$~Institute of Nuclear Research, 72 Tzarigradsko
chauss\'ee, 1784 Sofia, Bulgaria}
\EmailD{\href{mailto:dtrif@inrne.bas.bg}{dtrif@inrne.bas.bg}}

\ArticleDates{Received September 29, 2010, in f\/inal form December 04, 2010;  Published online December 15, 2010}

\Abstract{A new class of non-Hermitian Hamiltonians with real
spectrum, which are
written as a real linear combination of su(2) generators in the form $%
H=\omega J_{3}+\alpha J_{-}+\beta J_{+}$, $\alpha \neq \beta$, is
analyzed. The metrics which allows the transition to the equivalent
Hermitian Hamiltonian is established. A pseudo-Hermitian
supersymmetic extension of such Hamiltonians is performed. They
correspond to the pseudo-Hermitian supersymmetric systems of the
boson-phermion oscillators. We extend the supercoherent states
formalism to such supersymmetic systems via the pseudo-unitary
supersymmetric displacement operator method. The constructed
family of these supercoherent states consists of \textit{two dual subfamilies} that form a bi-overcomplete and bi-normal system in the
boson-phermion Fock space. The states of each subfamily are
eigenvectors of the boson annihilation operator and of one of the
two phermion lowering operators.}


\Classification{81Q12; 81Q60; 81R30}

\renewcommand{\thefootnote}{\arabic{footnote}}
\setcounter{footnote}{0}

\section{Introduction}

The study of non-Hermitian Hamiltonians with real spectrum has
received a great deal of interest during the last decade
\cite{Bender1,Mostafa02a}. One celebrated model of such
non-Hermitian PT-symmetric Hamiltonian is proposed by Swanson
\cite{Swanson}, which is expressed in terms of the usual harmonic
oscillator creation and annihilation operators $a^{\dagger }$ and
$a$, namely $H=\omega (a^{\dagger
}a+\frac{1}{2})+\alpha a^{2}+\beta a^{\dagger 2}$  with   $\omega $, $\alpha $, and $\beta $ real parameters, such that $\alpha
\neq \beta $ and $\omega ^{2}-4\alpha \beta >0$. This Hamiltonian
has been studied extensively in the literature by several authors
\cite{Swanson,Geyer04,Jones05,Bagchi2,Musumbu}. The metric operator
$\rho $, mapping $H$ to its Hermitian counterpart $h$ via the
relation $h=\rho H\rho ^{-1}$, has been constructed by using several
approaches. This Hamiltonian has been extended later on by Quesne
\cite{Quesne07,Quesne08} in the framework of the su(1,1) approach by
writing it as a linear combination of su(1,1) generators.

In the context of this extension, we introduce another kind of
non-Hermitian Hamiltonian with real spectrum in the Lie-algebric
framework \cite{Bagchi1,Paulo}, which is presented as a linear
combination of the generators $J_{-}$, $J_{+}$ and $J_{3}$ of the
su(2) Lie algebra. Then we introduce the pseudo-Hermitian
supersymmetric extension of such Hamiltonians, and naturally extend
the supercoherent states approach to such pseudo-Hermitian
supersymmetric systems. The Hermitian version of such Hamiltonians
has been widely used in the f\/ields of atomic physics and quantum
optics, in particular in the study of the interaction of two-level
atom systems with a coherent radiation f\/ield~\cite{Allen,Eberly,
Arecchi, Zhang}.

The organization of the paper is as follows. In Section~\ref{section2} we study our
pseudo-Hermitian Hamiltonian and we establish the metrics which
allows the transition to the corresponding Hermitian one. In Section~\ref{section3}
we consider a pseudo-Hermitian supersymmetric system in the form of
boson-phermion\footnote{The notion of pseudo-fermion (phermion) has
been introduced by Mostafazadeh~\cite{Mostafa04b}. The def\/ining algebra
of the phermions is a pseudo-Hermitian generalization of the
usual fermion algebra, namely: $\alpha ^{2}=\alpha ^{\#2}=0$,
$\alpha \alpha ^{\#}+\alpha ^{\#}\alpha =1,$ where $\alpha ^{\#} =
\eta ^{-1}\alpha ^{\dagger }\eta $ and $\alpha $ are respectively the creation
and annihilation operators of what was called the pseudo-Hermitian
fermion or simply a phermion~\cite{Mostafa04b}, and~$\eta $ is a linear, Hermitian,
and invertible operator. At $\eta =1$ one has $\alpha ^{\#}=\alpha^{\dagger }$,
i.e.\ the phermion algebra reduces to the usual fermion algebra.} oscillator.
In Section~\ref{section4} we construct the supercoherent states (SCS) from the
lowest (ground) eigenstates of the supersymmetric Hamiltonians $H_{s}$ and $%
H_{s}^{\dagger }$, by acting with \textit{a pair} of the
pseudo-unitary displacement operators. We show that these SCS are
eigenstates of the boson annihilation operator and of \textit{the pair}
of phermion annihilation operators. The set of such SCS form a
bi-normalized and bi-overcomplete system.  The paper ends
with concluding remarks.

\section{Non-Hermitian su(2) Hamiltonian}\label{section2}

We consider the following non-Hermitian Hamiltonian:
\begin{gather}
H=\omega (Y^{\dagger }Y-\tfrac{1}{2})+\alpha Y+\beta Y^{\dagger },
\label{H1}
\end{gather}%
where $\omega $, $\alpha $, and $\beta $ are real parameters such that $%
\alpha \neq \beta $ and $\omega ^{2}+4\alpha \beta >0$, $Y$ and
$Y^{\dagger } $ are fermion annihilation and creation operators
respectively, which obey the usual fermion algebra:
\begin{gather*}
\{Y,Y^{\dagger }\}\equiv YY^{\dagger }+Y^{\dagger }Y=1,\qquad
Y^{2}=Y^{\dagger }{}^{2}=0.  
\end{gather*}
The Hamiltonian \eqref{H1} is the non-Hermitian extension of the
Hermitian fermionic Hamiltonian studied in greater detail in our
recent work \cite{Cherbal10}. Due to the nilpotency of the fermionic
ope\-ra\-tors~$Y$ and~$Y^{\dagger }$, the Hamiltonian~\eqref{H1} which
is a linear combination of~$Y$,~$Y^{\dagger }$ and $(Y^{\dagger
}Y-\tfrac{1}{2})$, represents the most general form of such
non-Hermitian Hamiltonians. The three operators~$Y$,~$Y^{\dagger }$
and $(Y^{\dagger }Y-\tfrac{1}{2})$ close under commutation the su(2)
Lie algebra:
\begin{gather*}
\left[ J_{+},J_{-}\right] =2J_{3},\qquad \left[ J_{3},J_{\pm
}\right] =\pm J_{\pm },
\end{gather*}
where
\begin{gather*}
J_{+}=Y^{\dagger },\qquad J_{-}=Y,\qquad J_{3}=Y^{\dagger }Y-\tfrac{1}{2},
\end{gather*}
and $J_{3}^{\dagger }=J_{3}$,  $ J_{\pm }^{\dagger }=J_{\mp }.$ Thus
the Hamiltonian~(\ref{H1}) is expressed as:
\begin{gather}
H=\omega J_{3}+\alpha J_{-}+\beta J_{+}.  \label{H2}
\end{gather}%
It would be useful to mention that the Hermitian version of the
su(2) Hamiltonian \eqref{H2}, has been widely used in the f\/ields of
atomic and optical physics, and quantum optics, in the study of systems of
two-level atoms interacting resonantly with a coherent
radiation f\/ield \cite{Allen,Eberly,Arecchi,Zhang}. For this reason in the present paper
we are interesting in investigation of the supercoherent state formalism (in Section~\ref{section4}).

Following the procedure as in \cite{Quesne07,Quesne08}, the
non-Hermitian operator $H$ can be transformed into the
corresponding Hermitian Hamiltonian $h$ by means of the similarity
transformation
\begin{gather}
h=\rho H\rho ^{-1}.  \label{h1}
\end{gather}%
This means that $H$ admits positive-def\/inite metric operator $\eta
_{+}=\rho ^{2}$. We look for the mapping function $\rho $ in the
form
\begin{gather}
\rho =e^{\epsilon \left[ 2J_{3}+z(J_{-}+J_{+})\right] },
\label{rho1}
\end{gather}%
where $\epsilon $ and $z$ are real parameters.
By using 2$\times$2 matrix representation of $J_+$, $J_-$, and $J_3$,  namely
\begin{gather*}
J_{+}=\left(
\begin{array}{cc}
0 & 1 \\
0 & 0%
\end{array}%
\right) ,\qquad J_{-}=\left(
\begin{array}{cc}
0 & 0 \\
1 & 0%
\end{array}%
\right) ,\qquad J_{3}=\left(
\begin{array}{cc}
\frac{1}{2} & 0 \\
0 & -\frac{1}{2}%
\end{array}%
\right) ,
\end{gather*}
we f\/ind
\begin{gather}
\rho =e^{\epsilon \left[ 2J_{3}+z(J_{-}+J_{+})\right] }=\left(
\begin{array}{cc}
\cosh \theta +\epsilon (\sinh \theta )/\theta & \epsilon z(\sinh
\theta
)/\theta \\
\epsilon z(\sinh \theta )/\theta & \cosh \theta -\epsilon (\sinh
\theta
)/\theta%
\end{array}%
\right) ,  \label{rho2}
\end{gather}%
where $\theta =\epsilon \sqrt{1+z^{2}}$, and $\epsilon $ and $z$ are
related through formula
\begin{gather*}
\epsilon =\dfrac{1}{2\sqrt{1+z^{2}}}\,{\rm arctanh}\, \dfrac{(\alpha -\beta )%
\sqrt{1+z^{2}}}{\alpha +\beta -\omega z},\qquad z\in {\mathbb R}.
\end{gather*}
The mapping $\rho $ can also be written in the form
\begin{gather}
\rho =\left( \frac{\alpha +\beta -\omega z+(\alpha -\beta )\sqrt{1+z^{2}}}{\alpha +\beta -\omega z-(\alpha -\beta )\sqrt{1+z^{2}}}\right) ^{\frac{1}{4
\sqrt{1+z^{2}}}\left[ 2J_{3}+z(J_{-}+J_{+})\right] }.  \label{rho3}
\end{gather}%
Introducing (\ref{rho2}) into (\ref{h1}) we obtain the Hermitian $h$
in the form,
\begin{gather}
h=\delta J_{3}+\lambda (J_{-}+J_{+}),  \label{h2}
\end{gather}%
where $\delta $  and $\lambda $ are
given explicitly by
\begin{gather*}
\delta  = \frac{\omega +(\alpha +\beta )z-z(\alpha +\beta -\omega z)\sqrt{1-%
\tfrac{(\alpha -\beta )^{2}(1+z^{2})}{(\alpha +\beta -\omega z)^{2}}}}{%
1+z^{2}}, \\
 \lambda  = \frac{\omega z+(\alpha +\beta )z^{2}+(\alpha +\beta -\omega z)%
\sqrt{1-\tfrac{(\alpha -\beta )^{2}(1+z^{2})}{(\alpha +\beta -\omega z)^{2}}}%
}{2(1+z^{2})}.
\end{gather*}

It is worth noting that in terms of parameters $\epsilon$, $z$ and
group generators $J_i$ the above formulas are quite similar to the
corresponding ones for the case of su(1,1)~\cite{Quesne07,Quesne08}.
We would like however to emphasize that our $\theta =\epsilon
\sqrt{1+z^{2}}$ is manifestly real and positive, which means that in
the su(2) approach, the positivity of the Hermitian operator $\rho $
is ensured for any $z \in \mathbb{R}$, unlike the su(1,1) approach
case \cite{Quesne07,Quesne08}, where $z$ is restricted to the
interval $\left[ -1,1\right]$. This is the principal dif\/ference
between the metrics of the two approaches.

We note that formulas \eqref{rho1}, \eqref{rho3} for $\rho$ and
\eqref{H2}, \eqref{h1}, \eqref{h2} for $H$ and $h$ are valid in any
Hermitian representation of $J_i$. In the case of half integer spin
we can further express $h$ in terms of fermionic number operator,
and $H$~-- in terms of pseudo-Hermitian fermionic (phermionic) number
operator. In this aim we introduce the creation and annihilation
operators $b^{\dagger }$ and $b$ associated to the corresponding
Hermitian Hamiltonian $h$ given in equation~\eqref{h2} as,
\begin{gather*}
b=\frac{(\delta +\Omega)}{2\Omega}J_{-}+\frac{(\delta -\Omega)}{2\Omega}%
J_{+}-\frac{2\lambda }{\Omega}J_{3},  
\qquad
b^{\dagger }=\frac{(\delta -\Omega)}{2\Omega}J_{-}+\frac{(\delta +\Omega)}{%
2\Omega}J_{+}-\frac{2\lambda }{\Omega}J_{3},  
\end{gather*}
where
\begin{gather}
\Omega = \sqrt{\omega ^{2}+4\alpha \beta}.  \label{0214}
\end{gather}
The operators $b^{\dagger }$  and $b$ satisf\/ies the standard fermion
algebra:
\begin{gather*}
\{b,b^{\dagger }\}\equiv bb^{\dagger }+b^{\dagger }b=1,\qquad
b^{2}=b^{\dagger }{}^{2}=0.
\end{gather*}
In terms of $b$ and $b^{\dagger }$, the Hamiltonian $h$ is
factorized to the form of the fermionic oscillator,
\begin{gather}
 h=\Omega\big( b^{\dagger }b-\tfrac{1}{2}\big) ,  \label{h3}
\end{gather}
The number operators $N=b^{\dagger }b$ satisf\/ies
\begin{gather*}
\left[ b,N\right] =b,\qquad \big[ b^{\dagger },N\big] =-b^{\dagger },%
\qquad \big[ b,b^{\dagger }\big] =1-2N,
\end{gather*}
The Hilbert space of the single-fermion system is spanned by the two
eigenstates $\left\{ \left\vert 0\right\rangle,\, \left\vert
1\right\rangle \right\}$ of~$N$:
\begin{gather*}
b^{\dagger }b\left\vert n\right\rangle =n\left\vert n\right\rangle
,\qquad n=0,1.
\end{gather*}
The operators $b$ and $b^{\dagger }$ allow transitions between the
states as
\begin{gather*}
b\left\vert 0\right\rangle =0,\qquad b\left\vert 1\right\rangle
=\left\vert 0\right\rangle  ,\qquad b^{\dagger }\left\vert
1\right\rangle =0,\qquad b^{\dagger }|0\rangle =|1\rangle .
\end{gather*}
Now we can apply to $b$ and $b^{\dagger }$ a similarity transformation,
that is inverse to \eqref{h1} to get annihilation and creation operators $B$ and $%
B^{\#}$ associated to the quasi-Hermitian Hamiltonian~\eqref{H2},
\begin{gather}
B=\rho ^{-1}b\rho , \qquad B^{\#}=\rho ^{-1}b^{\dagger }\rho .
\label{0217}
\end{gather}
The operators $B$ and $B^{\#}$ are related via the pseudo-Hermitian
conjugation, $B^{\#} = \rho^{-2}B^{\dagger }\rho^{2}$ and satisf\/ies
the phermion algebra~\cite{Mostafa04b}
\begin{gather}
B^{2}=B^{\#2}=0,\qquad \big\{ B,B^{\#}\big\}
=BB^{\#}+B^{\#}B=1. \label{0333}
\end{gather}
Using then the equations \eqref{h1}, \eqref{h3} and \eqref{0217} we
get the pseudo-Hermitian Hamiltonian \eqref{H1} in a factorized form
\begin{gather}
H = \Omega\big( B^{\#}B-\tfrac{1}{2}\big) .  \label{H3}
\end{gather}
The phermionic ladder operators $B$ and $B^{\#}$ can be represented
in the form of (non-Hermitian) linear combination of the generators
$J_3$, $J_\pm$,
\begin{gather*}
B=\mu _{1}J_{-}+\mu _{2}J_{+}+2\mu _{3}J_{3},
\qquad
B^{\#}=\nu _{1}J_{-}+\nu _{2}J_{+}+2\nu _{3}J_{3},
\end{gather*}
  where $\mu _{1}$, $\mu _{2}$, $\mu _{3}$, $\nu _{1}$, $\nu _{2}$
and $\nu
_{3}$ are expressed in terms of the $H$- and $\rho$-parameters $\omega$, $\alpha$, $\beta$ and~$\epsilon$,~$z$ as follows
\begin{gather*}
\mu _{1}  = \frac{\delta +\Omega}{2\Omega}+\left[ \left( 1+\tau
+z^{2}\right) \epsilon \frac{\sinh \theta }{\theta }+\left( 1+\tau
\right)
\cosh \theta \right] \epsilon \frac{\sinh \theta }{\theta }, \\
\mu _{2}  = \frac{\delta -\Omega}{2\Omega}-\left[ \left( 1-\tau
+z^{2}\right) \epsilon \frac{\sinh \theta }{\theta }-\left( 1-\tau
\right)
\cosh \theta \right] \epsilon \frac{\sinh \theta }{\theta }, \\
\mu _{3}  = -\frac{\lambda }{\Omega}-\left[ \tau \epsilon
\frac{\sinh \theta
}{\theta }+\cosh \theta \right] z\epsilon \frac{\sinh \theta }{\theta }, \\
\nu _{1}  = \frac{\delta -\Omega}{2\Omega}-\left[ \left( 1-\tau
+z^{2}\right) \epsilon \frac{\sinh \theta }{\theta }+\left( 1-\tau
\right)
\cosh \theta \right] \epsilon \frac{\sinh \theta }{\theta }, \\
\nu _{2}  = \frac{\delta +\Omega}{2\Omega}+\left[ \left( 1+\tau
+z^{2}\right) \epsilon \frac{\sinh \theta }{\theta }-\left( 1+\tau
\right)
\cosh \theta \right] \epsilon \frac{\sinh \theta }{\theta }, \\
\nu _{3}  = -\frac{\lambda }{\Omega}-\left[ \tau \epsilon
\frac{\sinh \theta }{\theta }-\cosh \theta \right] z\epsilon
\frac{\sinh \theta }{\theta },
\end{gather*}
where $\tau =(\omega +(\alpha +\beta )z)/\Omega$.

Having analyzed the quasi-Hermitian Hamiltonian $H$ given in equations~\eqref{H1}, \eqref{H2}, we turn toward its pseudo-Hermitian
supersymmetric extension and to the construction of supercoherent
states for pseudo-Hermitian (supersymmetric) systems.

\section{Quasi-Hermitian supersymmetric extension}\label{section3}

Quantum-mechanical SUSY is extended to the case of general
pseudo-Hermitian Hamiltonians \cite{Mostafa02a,Mostafa04b,Mostafa02b,
Mostafa2004} by replacing the superalgebra of standard
SUSY quantum mechanics \cite{Krive,Cooper} by the
pseudo-superalgebra
\begin{gather}
Q^{2}=Q^{\#2}=0,\qquad \big\{ Q,Q^{\#}\big\} =2H_{s},  \label{025}
\end{gather}%
where all operators remain $\mathbb{Z}
_{2}$-graded as usual, the Hamiltonian $H_{s}$  is pseudo-Hermitian
with
respect to some $\mathbb{Z}_{2}$-graded  operator $\eta $: $H_{s}^{\dagger }=\eta H_{s}\eta ^{-1}$, $Q $ is the pseudo-Hermitian \mbox{(PH-SUSY)} gene\-ra\-tor (supercharge) and $Q^{\#}=\eta
^{-1}Q^{\dagger }\eta $  is the pseudo-adjoint of $Q$  with the same~$\eta $. Mostafazadeh has explored in~\cite{Mostafa04b} the
statistical origin of PH-SUSY quantum mechanics, showing that there
exist two types of PH-SUSY realizations. The f\/irst one corresponds
to exchange symmetry bet\-ween a boson and phermion; in this case the
metric operator is def\/inite and the phermions are physically
equivalent to the ordinary fermions. The second type, which is
fundamentally dif\/fe\-rent from the standard boson-fermion system,
corresponds to the exchange symmetry between a~boson and abnormal
phermion; in this case the metric operator is indef\/inite.

Since our Hamiltonian $H$ given in equation \eqref{H2} is
quasi-Hermitian, the supersymmetric extension corresponding to $H$
is characterized by the boson-phermion system described by the
following Hamiltonian \cite{Mostafa04b}:
\begin{gather}
H_{s}  = H_{b}+H   = \Omega\big(a^{\dagger }a+B^{\#}B\big),  \label{437}
\end{gather}
where $H_{b}=\Omega (a^{\dagger }a+\frac{1}{2})$ is the bosonic
contribution
and $H $ is the phermionic one, given in equation~\eqref{H3}; $%
\Omega$ is real and positive given in equation~\eqref{0214}, $%
a^{\dagger }$ and $a$ are the standard bosonic creation and
annihilation
operators ($[ a,a^{\dagger }] =\mathbf{1}$), and $B^{\#}$ and $B$
 are the phermionic creation and annihilation operators def\/ined by
the algebra given in equation~\eqref{0333}. The bosonic operators $a$ and
$a^{\dagger }$ are supposed \cite{Mostafa04b} to commute with any
phermionic operator constructed out of $B$, $B^{\#}$ and $\eta $:
\begin{gather}
\left[ a,B\right] =\big[ a,B^{\#}\big] =\left[ a,\eta \right] =0,
\nonumber\\
\big[ a^{\dagger },B\big] =\big[ a^{\dagger },B^{\#}\big]
=\big[ a^{\dagger },\eta \big] =0.  \label{029}
\end{gather}%
From the third relation in \eqref{029} in which $a^{\dagger }$
commute with~$\eta $, we have:
\begin{gather*}
a^{\#}=\eta ^{-1}a^{\dagger }\eta =\eta ^{-1}\eta a^{\dagger
}=a^{\dagger }.
\end{gather*}
Hence, for the bosonic operators $a^{\dagger }$ and $a$ the
pseudo-Hermitian conjugation operation $(^{\#})$ coincide with the
conjugation operation $(^{\dagger })$.

The equivalent Hermitian supersymmetric Hamiltonian is given by
\begin{gather*}
h_{s}  = \rho H_{s}\rho ^{-1}
  = \Omega (a^{\dagger }a+b^{\dagger }b)
\end{gather*}
The operator $h_{s}$ is in the form of boson-fermion oscillator
Hamiltonian. The supercharges $Q$  and $Q^{\#}$  associated to
$H_s$, equation~\eqref{437}, and satisfying the
equation \eqref{025} are given by
\begin{gather*}
Q=\sqrt{2\Omega} a^{\dagger }B,\qquad Q^{\#}=\sqrt{2\Omega} aB^{\#}.
\end{gather*}%
These $Q$ and $Q^{\#}$ commute with the Hamiltonian \eqref{437},
\begin{gather*}
\left[ Q,H_{s}\right] =0=\big[ Q^{\#},H_{s}\big] .
\end{gather*}%
Since $H_{s}$ is quasi-Hermitian with discrete spectrum, we can
introduce the complete bi-ortho\-nor\-mal eigenbasis $\{|\psi
_{(n,\epsilon )}\rangle ,|\phi _{(n,\epsilon )}\rangle \}$,
$n=0,1,2,3\dots$, and $\epsilon =0,1,$ associated to $H_{s}$, which
satisfy
\begin{gather*}
\langle \phi _{(n,\,\epsilon )}\left\vert \psi _{(m,\nu
)}\right\rangle =\delta _{nm}\delta _{\epsilon \nu },  
\\
\sum_{n}\sum_{\epsilon=0}^{1}\left\vert
\phi _{(n,\,\epsilon )}\right\rangle \langle \psi _{(n,\,\epsilon )}|=%
\sum_{n}\sum_{\epsilon=0}^{1}\left\vert
\psi _{(n,\,\epsilon )}\right\rangle \langle \phi _{(n,\,\epsilon )}|=%
\mathbf{1}.  
\end{gather*}
Using $h_{s}=\rho H_{s}\rho ^{-1}$ we can easily establish the
relations of the states $|\psi _{(n,\epsilon )}\rangle$, $|\phi
_{(n,\epsilon )}\rangle$ to the eigenstates $\left\vert n, \epsilon
\right\rangle$ of $h_{s}$,
\begin{gather*}
h_{s}\left\vert n, \epsilon \right\rangle =E_{n}\left\vert
n, \epsilon \right\rangle,
\end{gather*}
as follows:
\begin{gather*}
|\psi _{(n, \epsilon )}\rangle =\rho ^{-1}\left\vert n, \epsilon
\right\rangle  
\end{gather*}%
and
\begin{gather*}
|\phi _{(n, \epsilon )}\rangle =\rho \left\vert n, \epsilon
\right\rangle . 
\end{gather*}
Let us note that the structure of the Hilbert space of PH-SUSY
systems remain $\mathbb{Z}_{2}$-graded as in the usual SUSY, the
boson-phermion Fock space being $\mathcal{H}=\mathcal{H}_{B}\oplus
\mathcal{H}_{F}$ \cite{Mostafa04b,Mostafa2004}.

\section{Supercoherent states}\label{section4}

We embark now on the construction of the supercoherent states (SCS)
for our quasi-Hermitian SUSY Hamiltonian $H_{s}$ given in equation~\eqref{437}. We shall follow as close as possible the scheme of SCS
for SUSY Hamiltonians developed in papers
\cite{Aragone86,Fatyga91,Berube93} and the scheme of CS for the
pseudo-Hermitian Hamiltonians in \cite{Cherbal07,Trifonov09}\footnote{For more rigorous math treatment of the model see Bagarello's paper~\cite{Bagarello}.},
generalizing both of them to the PH-SUSY case. We give f\/irst the
representation space on which our SCS will be def\/ined. The set
$\{|\psi _{(n,\epsilon )}\rangle ,|\phi _{(n,\epsilon)}\rangle \}$
of the eigenstates of $H$ and $H^{\dagger }$ respectively spans the
Fock space on which our SCS will be def\/ined.

As in the case CS of phermion oscillator~\cite{Cherbal07} our SCS
are expected to take the form of two bi-overcomplete and
bi-normalized families,
this time in the boson-phermion Fock space $\mathcal{H}_{B}\oplus \mathcal{H}_{F} $. In this scheme we need to clarify f\/irst the action of boson
and phermion ladder operators on the eigenstates of the Hamiltonian
$H_{s}$ and its conjugate $H_{s}^{\dagger }$.
In  the set of eigenstates $|\psi _{(n, \epsilon )}\rangle $ of $H_{s}$, the bosonic
states correspond to $\epsilon = 0$,  while the
phermionic ones $|\psi _{(n,1)}\rangle $ correspond to $\epsilon =1$.
The boson operators $a$, $a^{\dagger}$ and the phermion operators $B$,
$B^{\#}$ act on the number states $|\psi _{(n,\epsilon )}\rangle $ as
follows (to be compared with the corresponding action of boson and
fermion ladder operators in the case of ordinary SUSY~\cite{Fatyga91}):
\begin{gather*}
a|\psi _{(n, \epsilon )}\rangle =\sqrt{n}|\psi _{(n-1, \epsilon
)}\rangle
,\qquad a^{\dagger }|\psi _{(n, \epsilon )}\rangle = \sqrt{n+1}
|\psi_{(n+1, \epsilon )}\rangle   ,
\\
B|\psi _{(n,\,0)}\rangle =0,\qquad B|\psi
_{(n, 1)}\rangle =|\psi _{(n, 0)}\rangle  ,
\\
B^{\#}|\psi _{(n, 1)}\rangle =0,\qquad B^{\#}|\psi
_{(n, 0)}\rangle =|\psi _{(n,1)}\rangle   .  
\end{gather*}%
The operator $B$ annihilates the bosonic states $|\psi
_{(n,0)}\rangle $, and $B^{\#}$ maps these state onto the
phermionic states $|\psi _{(n,1)}\rangle$. The boson-phermion ground
state is $|\psi _{(0,0)}\rangle $ which satisf\/ies the equations
\begin{gather}
a|\psi _{(0,0)}\rangle  = B|\psi _{(0,0)}\rangle =0,   \label{314} \\
Q|\psi _{(0,0)}\rangle  = Q^{\#}|\psi _{(0,0)}\rangle =0.\nonumber
\end{gather}
The operators $Q$ and $Q^{\#}$ act on the states $|\psi
_{(n,1)}\rangle $ and $|\psi _{(n,0)}\rangle $ as raising and
lowering operators:
\begin{gather*}
Q|\psi _{(n,1)}\rangle  = \sqrt{\Omega  (n+1)}|\psi _{(n+1,0)}\rangle ,
\qquad
Q^{\#}|\psi _{(n,0)}\rangle  = \sqrt{\Omega  n}|\psi
_{(n-1,1)}\rangle .
\end{gather*}
The operator $Q$ maps phermionic states onto bosonic ones, and
$Q^{\#}$ maps bosonic states onto phermionic ones.

After having introduced all the ingredients, we construct the SCS $%
\left\vert \alpha ,\xi \right\rangle $ associated to the Hamiltonian
\eqref{437} as the orbit of the ground state \eqref{314} under the
action of a \textit{pseudo-unitary} displacement operators $D(\alpha
,\xi )$ which realize a pseudo-Hermitian generalization of the
representation of the Heisenberg--Weyl super algebra, generated by
the boson and phermion operators $a$, $a^{\dagger }$, $B$, $B^{\#}$ and the
identity $\mathbf{1}$:
\begin{gather}
\left\vert \alpha ,\xi \right\rangle =D(\alpha ,\xi )|\psi
_{(0,0)}\rangle , \label{3.18}
\\
D(\alpha ,\xi )=e^{\left( \alpha a^{\dagger }-\alpha ^{\ast
}a+i\beta \mathbf{1+}B^{\#}\xi -\xi ^{\ast }B\right) },  \nonumber 
\end{gather}
where $\alpha $ is c-number, $\beta $ is real number, and $\xi $ is
complex Grassmann number \cite{Berezin1,Berezin2,Cahill99,Junker98}.
Let us recall that $\xi $ is nilpotent and anticommutes with its
conjugate,
\begin{gather}
\xi ^{2}=0,\qquad \xi ^{\ast 2}=0,\qquad \xi \xi ^{\ast }+\xi
^{\ast }\xi =0.  \label{3.20}
\end{gather}%
The integrations over $\xi $ and $\xi ^{\ast }$ are performed
according to the Berezin rules,
\begin{gather}
\int d\xi ^{\ast }d\xi \,\xi \xi ^{\ast }=1, \qquad \int d\xi ^{\ast }d\xi
\,\xi =\int d\xi ^{\ast }d\xi \,\xi ^{\ast }=\int d\xi ^{\ast }d\xi
\,1=0. \label{3.21}
\end{gather}%
As in the fermion case \cite{Cahill99} $\xi $'s commute with
ordinary complex numbers and boson operators, and anticommute with
phermion operators $B$  and  $B^{\#}$,
\begin{gather}
\left\{ \xi ,B\right\} =0,\qquad \left\{ \xi ^{\ast },B\right\} =0 ,\qquad \big\{ \xi ,B^{\#}\big\} =0,\qquad \big\{ \xi ^{\ast
},B^{\#}\big\} =0.  \label{3.22}
\end{gather}%
The pseudo-Hermitian conjugation reverses the order of all fermionic
quantities, both the operators and the Grassmann numbers:
\begin{gather}
\big(B^{\#}\xi +\xi ^{\ast }B\big)^{\#}=\xi ^{\ast }B+B^{\#}\xi  .
\label{3.23}
\end{gather}%
By using the Baker--Campbell--Hausdorf\/f formulas \cite{Gilmore74} the
displacement operators $D(\alpha ,\xi )$ is written (up to a constant phase factor)
in the form:
\begin{gather}
D(\alpha ,\xi )=e^{\big(-\frac{1}{2}\xi ^{\ast }\xi
-\frac{\left\vert \alpha \right\vert ^{2}}{2}\big)}e^{\alpha a^{\dagger
}}e^{-\xi B^{\#}}e^{-\alpha ^{\ast }a}e^{-\xi ^{\ast }B}.
\label{3.24}
\end{gather}%
The pseudo-Hermitian adjoint $D^{\#}(\alpha ,\xi )$ is given by
\begin{gather*}
D^{\#}(\alpha ,\xi )=e^{- ( \alpha a^{\dagger }-\alpha ^{\ast}a + B^{\#}\xi -\xi ^{\ast }B ) }.
\end{gather*}%
In the last expression of $D^{\#}(\alpha ,\xi )$, we have taken into
account that for all the bosonic operators, the pseudo-Hermitian
conjugation operation $(^{\#})$ coincides with the conjugation operation $%
(^{\dagger })$, which is expressed in the equation~\eqref{029} as
consequence of the fact that the bosonic operators~$a^{\dagger }$
and~$a$ commutes with~$\eta $ \cite{Mostafa04b}.

The displacement operator $D$ is \textit{pseudo-unitary}:  $%
D^{\#}D = \mathbf{1}=DD^{\#}$.
The substitution of the expression \eqref{3.24} of $D(\alpha ,\xi
)$ in the equation~\eqref{3.18} yields the following expression of SCS
$\left\vert \alpha ,\xi \right\rangle $,
\begin{gather}
\left\vert \alpha ,\xi \right\rangle =e^{-\frac{1}{2}\xi ^{\ast }\xi
}\left( \left\vert \alpha ,0\right\rangle -\xi \left\vert \alpha
,1\right\rangle \right).  \label{3.27}
\end{gather}%
Here $\left\vert \alpha ,0\right\rangle $ are the standard boson CS
(Glauber CS \cite{Glauber63}) given explicitly by
\begin{gather}
\left\vert \alpha ,0\right\rangle = e^{-\frac{\left\vert \alpha
\right\vert
^{2}}{2}}\sum_{n=0}^{\infty}\frac{\alpha ^{n}}{\sqrt{n!%
}}|\psi _{(n,0)}\rangle ,   \label{3.28}
\end{gather}
$|\psi _{(n,0)}\rangle \equiv \left\vert n\right\rangle $
representing the
Fock space for the standard bosonic harmonic oscillator. The states $%
\left\vert \alpha ,1\right\rangle $ are the phermionic states given
explicitly by
\begin{gather*}
\left\vert \alpha ,1\right\rangle =e^{-\frac{\left\vert \alpha
\right\vert
^{2}}{2}}\sum_{n=0}^{\infty}\frac{\alpha ^{n}}{\sqrt{n!%
}}|\psi _{(n,1)}\rangle .
\end{gather*}
  In the limit $\xi =0$, the expression \eqref{3.27} of
$\left\vert \alpha ,\xi \right\rangle $ recovers the standard boson
CS \eqref{3.28}, i.e.\ the PH-SUSY superalgebra is reduced to the
standard boson algebra.

The Hermitian adjoint of $\left\vert \alpha ,\xi \right\rangle $ is
\begin{gather*}
\left\langle \alpha ,\xi \right\vert =e^{-\frac{1}{2}\xi ^{\ast }\xi
}\left( \langle \alpha ,0| +\xi ^{\ast }\langle \alpha
,1|\right)  , 
\end{gather*}
and the inner product $\langle \alpha ,\xi |\alpha ,\xi \rangle \neq
1,$ which is due to the nonorthogonality of the states $|\psi
_{(n,\epsilon
)}\rangle $, the latter property being a consequence of the fact that $%
B^{\#}\neq B^{\dagger }$.

Now we have to examine for (over)completeness the set of $|\alpha
,\xi \rangle $. One can check (using the rules \eqref{3.20}--\eqref{3.23}) that the resolution of identity is not satisf\/ied,
because the states $|\psi _{(n, \epsilon )}\rangle $ do not form an
orthogonal basis, which is a consequence of the non-Hermiticity of the
Hamiltonian~\eqref{437}:
\begin{gather*}
\int |\alpha ,\xi \rangle \langle \alpha ,\xi |\,d\mu (\alpha )d\xi
^{\ast }d\xi  \neq \mathbf{1},\qquad d\mu (\alpha )=d\alpha
^{\ast }d\alpha /\pi .  
\end{gather*}
The useful way to solve this problem of the (over)completeness, is
to use the main idea introduced previously for the case of
pseudo-Hermitian CS in~\cite{Cherbal07,Trifonov09}, which consists
of introduction of a~complementary pair of lader operators, such
that the system of two complementary sets of CS forms the so-called
\textit{bi-orthonormal and bi-overcomplete system}. In this aim we
introduce the second dual ladder
operator $\tilde{B}$ which is associated to~$H^{\dagger }$. The $B$ and $%
\tilde{B}$  form a complementary pair of lader lowering operators.
$\tilde{B} $ is related to the annihilation operator~$b$ of~$h$ via
the similarity transformation:
\begin{gather*}
\tilde{B}=\rho b\rho ^{-1}.
\end{gather*}%
Its action on the eigenstates of $H^{\dagger }$ is
\begin{gather*}
\tilde{B}\left\vert \phi _{(n,0)}\right\rangle =0,\qquad
\tilde{B}\left\vert \phi _{(n,1)}\right\rangle =\left\vert \phi
_{(n,0)}\right\rangle  .
\end{gather*}%
The operator $\tilde{B}$ is nilpotent. The operator $B^{\dagger
}=\rho b^{\dagger }\rho ^{-1}$ is the creation operator for
$H^{\dagger }.$ In this
way one obtains a second pair of phermionic lowering and raising operators $%
\tilde{B}$ and $B^{\dagger }$,
\begin{gather}
\tilde{B}B^{\dagger }+B^{\dagger }\tilde{B}=1,\qquad
\tilde{B}^{2}=B^{\dagger 2}=0 ,  \label{3.40}
\\
B^{\dagger }\left\vert \phi _{(n,1)}\right\rangle =0,\qquad B^{\dagger }\left\vert \phi _{(n,0)}\right\rangle = \left\vert
\phi _{(n,1)}\right\rangle .\nonumber
\end{gather}%
In view of the phermionic algebra \eqref{3.40} we introduce new
displacement
super operators $\widetilde{D}(\alpha ,\xi ),$%
\begin{gather*}
\widetilde{D}(\alpha ,\xi )=e^{( \alpha a^{\dagger }-\alpha
^{\ast }a + B^{\dagger }\xi -\xi ^{\ast
}\tilde{B}) }.
\end{gather*}
We build up now the second family of SCS $\widetilde{|\xi ,\alpha
\rangle }$ according to the above described scheme (see equations~\eqref{3.18}, \eqref{3.27}),
\begin{gather}
\widetilde{|\alpha ,\xi \rangle }  =\widetilde{D}(\alpha ,\xi
)\left\vert
\phi _{(n,0)}\right\rangle   
   =e^{-\frac{1}{2}\xi ^{\ast }\xi }\big( \widetilde{|\alpha
,0\rangle }-\xi \widetilde{|\alpha ,1\rangle }\big) ,  \label{344}
\end{gather}
where $\left\vert \phi _{(n,0)}\right\rangle $ is the ground state of $%
H_{s}^{\dagger },$ $\widetilde{|\alpha ,0\rangle }$ and
$\widetilde{|\alpha ,1\rangle }$ being given explicitly by,
\begin{gather*}
\widetilde{|\alpha ,0\rangle }  = e^{-\frac{\left\vert \alpha
\right\vert
^{2}}{2}}\sum_{n=0}^{\infty}\frac{\alpha ^{n}}{\sqrt{n!%
}}\left\vert \phi _{(n,0)}\right\rangle ,  \qquad
\widetilde{|\alpha ,1\rangle }  = e^{-\frac{\left\vert \alpha
\right\vert
^{2}}{2}}\sum_{n=0}^{\infty}\frac{\alpha ^{n}}{\sqrt{n!%
}}\left\vert \phi _{(n,1)}\right\rangle .
\end{gather*}
Since the eigenstates $|\phi _{(n, \epsilon )}\rangle $ of
$H_{s}^{\dagger } $ are not orthogonal, the scalar product
$\widetilde{\langle \alpha ,\xi | }\widetilde{\alpha ,\xi \rangle
}$, like the previous one $\langle \alpha
,\xi | \alpha ,\xi \rangle$, is dif\/ferent from~1. The two subsets of states $%
\{\left\vert \alpha ,\xi \right\rangle\}$ and $\{ \widetilde{|\alpha
,\xi \rangle }\}$ are \textit{bi-normalized} instead:
\begin{gather*}
\widetilde{\langle \alpha ,\xi |}\alpha ,\xi \rangle =\langle \alpha
,\xi \widetilde{|\alpha ,\xi \rangle }=1.
\end{gather*}
By means of the two type of states $\left\vert \alpha ,\xi
\right\rangle $ and $\widetilde{|\alpha ,\xi \rangle }$ the
resolution of the identity is realized in the following way,
\begin{gather}
\int  |\alpha ,\xi \rangle \widetilde{\langle \alpha ,\xi |}d\mu
(\alpha )d\xi ^{\ast }d\xi =\int  \widetilde{|\alpha ,\xi \rangle
}\langle \alpha ,\xi |d\mu (\alpha )d\xi ^{\ast }d\xi
 =\mathbf{1}. \label{3.49}
\end{gather}
The equation \eqref{3.49} can be easily verif\/ied using the formulas
of $|\alpha ,\xi \rangle $ and $\widetilde{|\alpha ,\xi \rangle }$
(equations~\eqref{3.27} and \eqref{344}) and the
rules of integration \eqref{3.21}. Thus the system of states $%
\{\left\vert \alpha ,\xi \right\rangle ,\widetilde{|\alpha ,\xi
\rangle }\}$ is bi-overcomplete in boson-phermion Fock space. It is
this system that we call \textit{boson-phermion SCS}, or more
shortly \textit{pseudo-Hermitian SCS}.

We would like to emphasize that these SCS satisfy also the f\/irst
def\/inition of usual coherent states (CS) given by Glauber
\cite{Glauber63} as eigenstates of boson and phermion annihilation
operators:
\begin{gather*}
a|\alpha ,\xi \rangle =\alpha |\alpha ,\xi \rangle
,\qquad
B|\alpha ,\xi \rangle =\xi |\alpha ,\xi \rangle , \nonumber\\
  a\widetilde{|\alpha ,\xi \rangle }=\alpha
\widetilde{|\alpha ,\xi \rangle },\qquad \tilde{B}\widetilde{|\alpha
,\xi \rangle }=\xi
\widetilde{|\alpha ,\xi \rangle } .
\end{gather*}
Finally, in the limit of $\eta =1$ (that is $B^{\#}\equiv B^{\dagger
}$), our SCS recover the standard SCS for SUSY boson-fermion
oscillator~\cite{Fatyga91,Berube93}. In the double
limits of $\eta =1$ and $\xi =0,$ these SCS coincide with of the
standard Glauber's CS \cite{Glauber63}.

\section{Concluding remarks}\label{section5}

In this paper, we have achieved some extensions in the framework of
the pseudo-Hermitian quantum mechanics. We have extended the study
of the Hermitian su(2) Hamiltonians to the case of non-Hermitian
su(2) Hamiltonians with real spectrum. For such pseudo-Hermitian
Hamiltonian system we established the metrics which allows the
transition to the corresponding Hermitian Hamiltonian. The
constructed metric operator depends on one real parameter as in the
case of the su(1,1) approach, this time however the real parameter
is not restricted by any inequality.

We have also extended the supercoherent states (SCS) approach to
pseudo-Hermitian supersymmetric (PH-SUSY) system characterized by
the boson-phermion oscillator \cite{Mostafa04b,Mostafa2004}. The
supersymmetric displacement operator method and the ladder operator
method for construction of ordinary SCS~\cite{Fatyga91,Berube93} are
both extended to the case PH-SUSY systems. For the boson-phermion
systems there are two complimentary pairs of ladder operators which
have to be used to construct bi-orthonormal Fock states and coherent
states. As a result the set of constructed SCS for the
boson-phermion system consists of a pair of two (dual) subsets of states
which are bi-normal and form a bi-overcomplete system in the
corresponding Hilbert space. The states of each subset are
eigenvectors of the boson annihilation operator and of the
corresponding phermionic lowering operator. In the limit of
Hermitian SUSY system our states recover the known SCS
\cite{Fatyga91,Berube93}.

The  PH-SUSY and pseudo-SCS scheme, developed in this paper for the
boson-phermion systems, can be naturally extended and applied to study
other (more general) PH-SUSY systems, such as system of pseudo-bosons
\cite{Trifonov09} and fermions  and system of pseudo-bosons and phermions.

\pdfbookmark[1]{References}{ref}
\LastPageEnding

\end{document}